\documentclass{article}
\usepackage{spconf,amsmath,graphicx,hyperref}

\usepackage{wrapfig}
\usepackage{graphicx}
\usepackage{subfigure}
\usepackage{multirow, makecell}
\usepackage{enumitem}
\usepackage{booktabs}
\usepackage{amsfonts}
\usepackage{cite}
\usepackage[capitalize,noabbrev]{cleveref}

\title{FocalCodec-Stream: \\ Streaming Low-Bitrate Speech Coding via Causal Distillation}
\name{Luca Della Libera$^{1,2}$, Cem Subakan$^{3,1,2}$, Mirco Ravanelli$^{1,2}$
}
\address{\textit{$^1$Concordia University, 
  $^2$Mila-Quebec AI Institute,
  $^3$Université Laval}
}

\begin{document}
\ninept
\maketitle
\begin{abstract}
Neural audio codecs are a fundamental component of modern generative audio pipelines. Although recent codecs achieve strong low-bitrate reconstruction and provide powerful representations for downstream tasks, most are non-streamable, limiting their use in real-time applications.
We present FocalCodec-Stream, a hybrid codec based on focal modulation that compresses speech into a single binary codebook at 0.55 - 0.80 kbps with a theoretical latency of 80 ms. Our approach combines multi-stage causal distillation of WavLM with targeted architectural improvements, including a lightweight refiner module that enhances quality under latency constraints. Experiments show that FocalCodec-Stream outperforms existing streamable codecs at comparable bitrates, while preserving both semantic and acoustic information. The result is a favorable trade-off between reconstruction quality, downstream task performance, latency, and efficiency. Code and checkpoints will be released at \href{https://github.com/lucadellalib/focalcodec}{https://github.com/lucadellalib/focalcodec}.
\end{abstract}
\begin{keywords}
Speech coding, discrete tokens, streamability
\end{keywords}

\section{Introduction}
\label{sec:intro}
Audio coding is a fundamental signal processing technique that compresses speech into compact discrete representations at low bitrates and reconstructs them with minimal perceptual loss. Originally developed for efficient transmission to reduce network traffic~\cite{zeghidour2021soundstream, defossez2023encodec}, neural audio codecs (NACs) have advanced rapidly in recent years and have become central to modern speech generative models~\cite{mousavi2025survey}.
This shift is largely driven by the success of large language models~\cite{openai2023gpt4, chowdhery2024palm, jiang2024mixtralexperts, dubey2024llama3herdmodels}, which established autoregressive sequence modeling as the dominant paradigm in text and are now being extended to other modalities, including audio.

In this context, NACs provide \textbf{discrete audio tokens} enabling applications such as text-conditioned audio generation~\cite{borsos2023audiolm, copet2023musicgen, kreuk2023audiogen, wang2023valle} and speech language models (SLMs)~\cite{zhang2023speechgpt, hassid2023textually, defossez2024moshi, nguyen2024spiritlminterleavedspokenwritten}. To support these tasks, codec tokens should preserve both acoustic and semantic information while maintaining high reconstruction quality. Additionally, bitrate and sequence length become especially critical: unlike transmission, where even codecs at 12 kbps already outperform traditional codecs~\cite{defossez2023encodec}, much lower bitrates are required for training effective SLMs~\cite{defossez2024moshi}.

Recent work has sought to address these challenges. Hybrid codecs enrich acoustic tokens with semantic content~\cite{zhang2024speechtokenizer, defossez2024moshi}, some employ supervised fine-tuning~\cite{parker2024scaling}, others reduce frame rate~\cite{defossez2024moshi, parker2024scaling,casanova2025nano}, and a growing trend compresses all the information into a single codebook~\cite{ji2024wavtokenizer, xin2024bigcodec, wu2024ts3codec, dellalibera2025focalcodec, ye2025llasa}. 
A remaining challenge, however, is \textbf{streamability}: most codecs are designed for offline processing and depend on long future context windows, making them unsuitable for real-time tasks such as speech assistants, interactive dialogue, and low-latency generation. 
While some codecs~\cite{defossez2023encodec, wu2023audiodec, ahn2024hilcodec, defossez2024moshi, wu2024ts3codec, parker2024scaling, hartuv2025past} already support streaming, they typically compromise on other aspects, such as requiring high bitrates~\cite{defossez2023encodec, wu2023audiodec, ahn2024hilcodec}, relying on multiple codebooks~\cite{defossez2023encodec, wu2023audiodec, ahn2024hilcodec, defossez2024moshi, parker2024scaling, hartuv2025past}, neglecting semantic information~\cite{defossez2023encodec, wu2023audiodec, ahn2024hilcodec, wu2024ts3codec}, or providing weak representations for downstream tasks~\cite{defossez2023encodec, wu2023audiodec, ahn2024hilcodec}.  
This highlights the need for low-bitrate, single-codebook codecs that unify semantic and acoustic representations, maintain high reconstruction quality, and support streamability. 

In our prior work~\cite{dellalibera2025focalcodec}, we introduced FocalCodec, an efficient single-codebook codec based on WavLM~\cite{chen2022wavlm}, focal modulation~\cite{yang2022focalnets, dellalibera2024focal}, and binary spherical quantization~\cite{zhao2024bsq}, which satisfies all requirements except streamability. In this paper, we extend FocalCodec to the streaming setting.
In particular, our contributions are as follows:
\begin{itemize}[topsep=0.2cm, itemsep=0cm, leftmargin=0.35cm]
    \item We introduce {FocalCodec-Stream}, a novel \textbf{hybrid} codec that compresses speech into a \textbf{single binary codebook at low bitrates} (0.55 - 0.80 kbps) while supporting \textbf{streaming} inference with a theoretical latency of 80 ms.
    \item To achieve this, we propose a \textbf{multi-stage causal distillation} strategy to adapt WavLM for streaming, combined with targeted architectural modifications. In particular, we design a lightweight \textbf{refiner} module that mitigates quality degradation under latency constraints.
    \item We evaluate FocalCodec-Stream extensively across \textbf{reconstruction and downstream tasks}, and validate our design through {ablation studies}.
\end{itemize}

\begin{figure*}[t!]
  \centering
\includegraphics[width=0.95\textwidth]{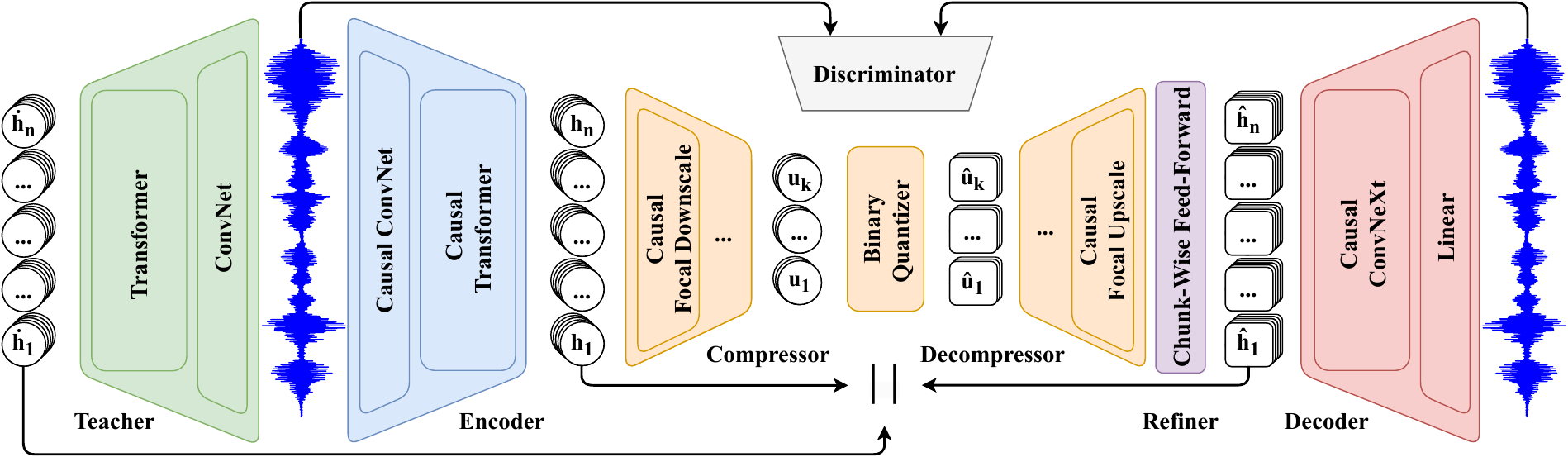}
  \vspace{-0.2cm}
  \caption{FocalCodec-Stream architecture. The encoder extracts features containing both acoustic and semantic information. These features are then mapped to a low-dimensional space by the compressor, binary quantized, and projected back by the decompressor. The decoder resynthesizes the waveform from these features. All these modules are causal, while a non-causal teacher is used for distillation to align intermediate causal features with their non-causal counterparts.}
  \label{fig:focalcodec_causal}
\vskip -0.4cm
\end{figure*}

\section{FocalCodec-Stream}
\label{sec:method}

\subsection{Codec architecture}
Our streaming codec builds upon the original FocalCodec~\cite{dellalibera2025focalcodec} architecture, with several modifications to enable low-latency streaming inference (see \cref{fig:focalcodec_causal}). We target a latency of 80 ms, which has been shown sufficient for building highly responsive SLMs~\cite{defossez2024moshi}.

The \textbf{encoder}, consisting of the first six layers of \href{https://github.com/microsoft/unilm/tree/master/wavlm}{WavLM-large}, is made streamable by replacing standard convolutions with causal convolutions and full-context gated relative attention with sliding window gated relative chunked attention, with chunks of 80 ms (4 feature frames). Compared to strictly causal attention, the sliding window approach discards old context to ensure constant memory usage and enable infinite streaming, while chunking introduces a lookahead of up to 3 frames, improving performance within the latency budget. Unlike pooling-based frame reduction, this design preserves WavLM compatibility without requiring full retraining from scratch.
The \textbf{compressor} and \textbf{decompressor}, built on focal modulation~\cite{yang2022focalnets,dellalibera2024focal}, combine depth-wise convolutions, position-wise feed-forward layers, nonlinearities, and global pooling. For streaming, we replace standard convolutions with causal convolutions and substitute global pooling with a large-kernel causal convolution. This acts as a learnable sliding window moving average.
The binary spherical \textbf{quantizer}~\cite{zhao2024bsq} operates independently on each latent and is naturally streamable.
The Vocos~\cite{siuzdak2023vocos} \textbf{decoder} is made streamable by replacing the ConvNeXt blocks with causal convolutions and substituting the inverse STFT with a linear projection and flattening, following~\cite{okamoto2023wavenext}: each hidden state is projected to $K$ entries, where $K$ is the upsampling factor, and then flattened to reconstruct the waveform.
Note that, except for the chunked attention, most modules are strictly causal, incurring just 20 ms latency instead of the overall 80 ms latency budget.

To address this, we introduce a lightweight \textbf{refiner} module after the decompressor. Implemented as a residual chunk-wise feed-forward layer, it leverages the available latency to better align with WavLM features while remaining efficient.
Let $x \in \mathbb{R}^{N \times D}$ where $N$ is the sequence length and $D$ the hidden size. We first reshape it into chunks $x_c \in \mathbb{R}^{{N}/{C} \times C D}$ of size $C$. For each chunk vector $x_c$: 
\vspace{-3pt}
\begin{equation}
\hat{x}_c \;=\; x_c \;+\; W_{\text{out}} \,\mathrm{GELU}\!\left(W_{\text{in}} x_c + b_{\text{in}}\right) + b_{\text{out}},
\vspace{-2pt}
\end{equation}
where $W_{\text{in}}, W_{\text{out}} \in \mathbb{R}^{C D \times C D}$ and $b_{\text{in}}, b_{\text{out}} \in \mathbb{R}^{C D}$. Finally, we unflatten $\hat{x}_c$ back to the original shape $\hat{x} \in \mathbb{R}^{N \times D}$.
Empirically, we find the refiner useful for improving perceptual quality at the target latency, while having minimal impact on inference speed.

\subsection{Causal distillation}
The main challenge in developing a streaming version of FocalCodec lies in making the full-context WavLM encoder streamable. Prior work has explored ad-hoc self-supervised approaches~\cite{chiu2022bestrq}, and more recently proposed methods for adapting off-the-shelf self-supervised models~\cite{fu2024wav2vecstreaming} and discrete units~\cite{choi2025stream}. However, these efforts have been largely limited to speech recognition and focus primarily on semantic information. Their impact on speech resynthesis quality remains unclear, as does the ability to preserve strong representations for downstream tasks. Notably, \cite{meng2025context} shows that the effective receptive field of self-supervised speech models is relatively small in deeper layers, suggesting that these layers can be adapted more readily for streaming, whereas earlier layers depend on much broader temporal context.
To overcome these challenges, we propose a {multi-stage distillation} framework, where the original full-context WavLM encoder serves as the teacher:

\begin{itemize}[topsep=0.1cm, itemsep=0cm, leftmargin=0.35cm]

\item \textbf{Stage-1.} We first assess the contribution of different architectural modules in the original FocalCodec@50 model under streaming constraints, where only the encoder is converted to streaming while all downstream components remain unchanged. This analysis shows that the most critical components are (1) the learned positional embedding and (2) the full-context gated relative attention, whereas the convolutional feature extractor plays a comparatively minor role. In particular, the positional embedding relies on a convolution with a receptive field of 2.56 s (128 feature frames), which far exceeds the 80 ms latency budget and would in practice impose an initial delay of seconds before playback can begin. To address this, we causally distill the positional embedding by encouraging its causal variant to approximate the full-context one, minimizing an L2 loss on training data. This procedure is lightweight, requiring only a single convolutional layer with 8.4M parameters (less than 10\% of the encoder size).

\item \textbf{Stage-2.} After distilling the positional embedding, we proceed to distill the attention and convolutional feature extractor. These components are converted to causal form, then distilled from the full-context teacher. To guide the process, we minimize an L2 loss between the causal and teacher embeddings at each of the six layers, applied pairwise. Since the final layers are most important, we weight their losses more heavily using a reversed linear schedule: 1.0 for layer 6, 0.9 for layer 5, 0.8 for layer 4 etc. This gives the earlier layers more freedom to adapt while preserving the representational strength of the deeper layers.

\item \textbf{Stage-3.} Once the WavLM encoder has been distilled, we train the causal compressor–quantizer–decompressor system on top of it, following the same recipe as~\cite{dellalibera2025focalcodec}. We use an L2 loss between the causally distilled and decompressed representations. In parallel, we train the causal decoder using the original {full-context} WavLM representations.

\item \textbf{Stage-4.} A distribution shift remains between the decoder trained on full-context features and the decompressor trained to reconstruct causally distilled features. To close this gap, we introduce the learnable refiner module described earlier. During this stage, we fine-tune the encoder, compressor, quantizer, decompressor, and refiner jointly, while applying an L2 loss between the teacher output and the refiner output. This provides the system with additional capacity to adapt, allowing the encoder to generate features that are not exact copies of the teacher's but are more suitable for reconstruction by the decompressor. Although this step is optional, since the codec is already functional after the previous stage, we find it considerably improves reconstruction quality.
\end{itemize}

\section{Experimental Setup}
\label{sec:setup}
We follow the hyperparameter settings of \cite{dellalibera2025focalcodec}, with modifications to compensate for the expected performance degradation introduced by causality. Specifically, we increase the capacity of the compressor and decompressor by setting the hidden size of each of the three focal downscaling/upscaling layers to 1024, and by doubling both the focal window and focal factor to 14 and 4, respectively. In addition, we replace layer normalization with the lightweight DyT nonlinearity~\cite{zhu2025dyt}. For all sliding-window operations, the past context is set to 10.24 s (512 feature frames).
For the decoder, we increase the hidden size to 1024 and the feed-forward dimension to 2048. Importantly, we train with an upsampling factor of 480 instead of 320, enabling the model to reconstruct audio at 24 kHz rather than 16 kHz. This provides the decoder with \textbf{super-resolution} capabilities, which we find beneficial for enhancing audio quality.
We focus on the 50 Hz version and train three variants with codebook sizes of 2048, 4096, and 65,536.

To accommodate the increasing model capacity, we progressively scale the training data across the distillation stages. Stage-1 uses LibriTTS~\cite{zen2019libritts} (585 hours), resampled to 16 kHz. Stage-2 and stage-3 use Libri-Light-medium~\cite{kahn2020libri} (5k hours), trained on fixed-length chunks of 15 s. Stage-4 scales to the full Libri-Light corpus (60k hours) while increasing the chunk size to 30 s.
For the decoder, we follow the original training recipe and use only the \texttt{clean-100} split of LibriTTS at its original 24 kHz sampling rate, with chunk size set to 3 s. All models are trained on NVIDIA A100 (80 GB) GPUs with a learning rate that decreases when validation loss fails to improve for a few epochs, and training is stopped once no improvement is observed for several consecutive epochs. In some runs, we found it useful to reset the learning rate once the validation loss appeared to have converged, which yielded a few additional improvements. Implementation details and hyperparameters will be provided in the accompanying code repository.

\section{Results}
\label{sec:results}
We compare our models against recent streaming low-bitrate codecs with publicly available checkpoints, excluding works such as \cite{parker2024scaling} and \cite{wu2024ts3codec}, which do not provide streaming checkpoints.
Since our focus is on the low-bitrate regime, whenever multiple quantizer configurations are available, we set them to achieve a bitrate close to the range of 0.55 - 0.80 kbps, ensuring a fair comparison. In particular, we include EnCodec~\cite{defossez2023encodec}, AudioDec~\cite{wu2023audiodec}, and HILCodec~\cite{ahn2024hilcodec}, which are inherently acoustic codecs, as well as Mimi~\cite{defossez2024moshi} and PAST~\cite{hartuv2025past}, which are hybrid approaches based on distillation and supervised fine-tuning, respectively. We also include the original non-streaming FocalCodec@50~\cite{dellalibera2025focalcodec} for comparison. The configurations and details of each model are summarized in \cref{tab:baselines}.

\subsection{Speech resynthesis and voice conversion}
First, we evaluate FocalCodec-Stream on speech resynthesis (\textbf{SR}), considering both \textbf{English} and \textbf{multilingual} speech. We follow the same experimental protocol as in \cite{dellalibera2025focalcodec}. For English speech, we use the {LibriSpeech}~\cite{panayotov2015librispeech} \texttt{test-clean}, while for multilingual speech, we use a subset of MLS~\cite{pratap2020mls}.
We report UTMOS~\cite{saeki2022utmos} for naturalness, Whisper-small~\cite{radford2022robust} dWER for intelligibility, WavLM-based embedding similarity for speaker fidelity, code usage and normalized entropy for codebook efficiency, and real-time factor (RTF) for inference speed, measured on a 1/8 NVIDIA H100 (80 GB) multi-instance GPU partition.
\cref{tab:speech_resynthesis} shows the results. On LibriSpeech, FocalCodec-Stream achieves UTMOS close to full-context FocalCodec and outperforms all baselines in intelligibility, with FocalCodec-S@50-65k reaching the best dWER and highest speaker similarity. Code usage is 100\%  with high entropy, indicating efficient utilization of the quantizer. On MLS, performance drops across all models due to multilingual variability, but FocalCodec-Stream variants still outperform acoustic codecs and remain competitive with Mimi6. In particular, FocalCodec-S@50-65k shows the best trade-off between intelligibility and speaker fidelity. By contrast, PAST, while achieving good dWER on LibriSpeech, performs poorly in this setting because it was fine-tuned in a supervised fashion on English data only and thus fails to generalize to multilingual audio. This highlights a key limitation of supervised adaptation approaches, which are prone to overfitting to domain-specific data.

We also perform one-shot voice conversion (\textbf{VC}) experiments to assess the ability of FocalCodec-Stream to disentangle content from speaker information. Following the protocol of \cite{dellalibera2025focalcodec}, we use a dataset of parallel utterances derived from {VCTK}~\cite{veaux2017cstr}. FocalCodec-Stream consistently achieves substantially better naturalness and intelligibility than prior streaming codecs. Most notably, FocalCodec-S@50-65k attains both the highest UTMOS and speaker similarity, while keeping dWER significantly lower than acoustic codecs. Compared to Mimi6, which achieves competitive similarity, FocalCodec-S maintains much lower dWER. Conversely, although PAST reports slightly better dWER, it performs substantially worse in terms of speaker similarity. This highlights that FocalCodec-S is the only streaming codec to simultaneously achieve high intelligibility and strong speaker fidelity in this setting, both of which are equally critical for successful voice conversion.

\begin{table}[t!]
\setlength{\tabcolsep}{3pt}
\vspace{-6pt}
\caption{Codecs considered in our experiments.}
\label{tab:baselines}
\centering
\resizebox{0.45\textwidth}{!}{%
\begin{tabular}{lccccccc}
\toprule
\textbf{Codec} & \makecell{\textbf{Bitrate} \\ \textbf{(kbps)}} & \makecell{\textbf{Sample} \\ \textbf{Rate} \\ \textbf{(kHz)}} & \makecell{\textbf{Token} \\ \textbf{Rate} \\ \textbf{(Hz)}} & \makecell{\textbf{Codebooks}} & \makecell{\textbf{Latency} \\ \textbf{(ms)}} & \makecell{\textbf{Params} \\ \textbf{(M)}} \\
\midrule
EnCodec  & 1.50 & 24 & 75.0  & 2 $\times$ 1024 & 13 & 15  \\
AudioDec & 1.60 & 24 & 80.0  & 2 $\times$ 1024 & 13 & 8   \\
HILCodec & 1.50 & 24 & 75.0  & 2 $\times$ 1024 & 13 & 11  \\
Mimi5    & 0.69 & 24 & 12.5  & 5 $\times$ 2048 & 80 & 82  \\
Mimi6    & 0.83 & 24 & 12.5  & 6 $\times$ 2048 & 80 & 82  \\
PAST     & 1.00 & 16 & 50.0  & 2 $\times$ 1024 & 20 & 126 \\
\midrule
\textbf{FocalCodec-S@50-2k}  & 0.55 & 16 / 24 & 50.0  & 1 $\times$ 2048 & 80 & 249 \\
\textbf{FocalCodec-S@50-4k}  & 0.60 & 16 / 24 & 50.0  & 1 $\times$ 4096 & 80 & 249 \\
\textbf{FocalCodec-S@50-65k} & 0.80 & 16 / 24 & 50.5  & 1 $\times$ 65536 & 80 & 249 \\
\midrule
FocalCodec@50  & 0.65 & 16 & 50.0 & 1 $\times$ 8192 & --- & 142 \\
\bottomrule
\end{tabular}
}
\vskip -0.2in
\end{table}

\begin{table*}[t!]
\setlength{\tabcolsep}{2pt} 
\vspace{-6pt}
\caption{Speech resynthesis and voice conversion. \textbf{Best}, \underline{second-best} and \setlength{\fboxsep}{1pt}\fbox{{best non-streaming}} results are highlighted.}
\label{tab:speech_resynthesis}
\begin{center}
\begin{footnotesize}
\resizebox{0.95\textwidth}{!}{%
\begin{tabular}{lc|cccccc|cccccc|ccc}
\toprule
\multirow{4}{*}{\textbf{Codec}} & \multirow{4}{*}{\makecell{\textbf{Bitrate} \\ \textbf{(kbps)} } $\downarrow$} 
& \multicolumn{6}{c|}{\textbf{SR -- English}} 
& \multicolumn{6}{c|}{\textbf{SR -- Multilingual}}
& \multicolumn{3}{c}{\textbf{VC}} \\
\cmidrule{3-17}
&
& \textbf{UTMOS} $\uparrow$ & \textbf{dWER} $\downarrow$ & \textbf{Sim} $\uparrow$ & \makecell{\textbf{Code} \\ \textbf{Usage}} $\uparrow$ & \makecell{\textbf{Norm.} \\ \textbf{Entropy}} $\uparrow$ & \textbf{RTF} $\uparrow$
& \textbf{UTMOS} $\uparrow$ & \textbf{dWER} $\downarrow$ & \textbf{Sim} $\uparrow$ & \makecell{\textbf{Code} \\ \textbf{Usage}} $\uparrow$ & \makecell{\textbf{Norm.} \\ \textbf{Entropy}} $\uparrow$ & \textbf{RTF} $\uparrow$
& \textbf{UTMOS} $\uparrow$ & \textbf{dWER} $\downarrow$ & \textbf{Sim} $\uparrow$ \\
\midrule
\multicolumn{1}{l}{Reference} & ---  
& 4.09 & 0.00 & 100.0 & --- & --- & --- 
& 2.84 & 0.00 & 100.0 & --- & --- & --- 
& 4.09 & 0.00 & 100.0 \\
\multicolumn{1}{l}{EnCodec} & 1.50 
& 1.58 & 8.08 & 93.8  & 93.4 & 82.1 & 91 
& 1.33 & 29.60 & 95.5  & 93.4 & 79.2 & 113 
& 1.24 & 86.52 & 72.2  \\
\multicolumn{1}{l}{AudioDec} & 1.60 
& 1.48 & 11.61 & 92.1 & 91.9 & 70.0 & 145 
& 1.29 & 40.95 & 92.3 & 87.5 & 68.2 & 195 
& 1.26 & 68.45 & 68.2 \\
\multicolumn{1}{l}{HILCodec} & 1.50 
& 2.86 & 6.65 & 95.4 & \underline{99.0} & 95.6 & 41 
& 1.81 & 25.32 & \underline{97.8} & 99.1 & 94.8 & 41
& 1.40 & 58.36 & 76.8 \\
\multicolumn{1}{l}{Mimi5} & 0.69 &
3.29 & 5.73 & 96.0  & 95.6 & 91.8 & \textbf{157} &
2.08 & 30.96 & 96.7  & 95.9 & 89.0 & \textbf{219}
& 2.40 & 110.00 & 89.7 \\
\multicolumn{1}{l}{Mimi6} & 0.83 
& 3.44 & 4.77 & \underline{96.6} & 96.2 & 92.0 & \underline{154} 
& 2.19 & 26.12 & 97.4 & 96.5 & 89.2 & \underline{216} 
& 2.62 & 110.00 & 91.3 \\
\multicolumn{1}{l}{PAST} & 1.00 
& 2.33 & \underline{4.04} & 83.8 & 56.7 & 90.7 & 59 
& 1.44 & 49.35 & 80.8 & 57.0 & 87.5 & 63
& 1.42 & \textbf{18.28} & 68.5 \\
\midrule
\multicolumn{1}{l}{\textbf{FocalCodec-S@50-2k}} & \textbf{0.55} 
& \textbf{3.88} & 4.63 & 96.1 & \textbf{100.0} & \textbf{99.4} & 106 
& \textbf{2.68} & 24.64 & 97.5 & \textbf{100.0} & \underline{98.8} & 108 
& \underline{2.72} & 25.08 & \underline{92.1} \\
\multicolumn{1}{l}{\textbf{FocalCodec-S@50-4k}} & \underline{0.60} 
& \underline{3.87} & 4.39 & 96.3 & \textbf{100.0} & \underline{99.1} & 106 
& \textbf{2.68} & \underline{23.69} & 97.6 & \textbf{100.0} & \textbf{98.9} & 108 
& \underline{2.72} & 24.39 & 91.5 \\
\multicolumn{1}{l}{\textbf{FocalCodec-S@50-65k}} & 0.80 
& 3.85 & \textbf{3.68} & \textbf{97.0} & \textbf{100.0} & 98.7 & 106 
& \underline{2.65} & \textbf{19.88} & \textbf{98.1} & \underline{99.2} & 98.3 & 107 
& \textbf{3.10} & \underline{22.71} & \textbf{92.5} \\
\midrule
\multicolumn{1}{l}{FocalCodec@50} & 0.65 
& \setlength{\fboxsep}{1pt}\fbox{{4.05}} & \setlength{\fboxsep}{1pt}\fbox{{2.18}} & \setlength{\fboxsep}{1pt}\fbox{{97.4}} & \setlength{\fboxsep}{1pt}\fbox{{100.0}} & 98.9 & 123 
& \setlength{\fboxsep}{1pt}\fbox{{2.96}} & \setlength{\fboxsep}{1pt}\fbox{{12.57}} & \setlength{\fboxsep}{1pt}\fbox{{98.3}} & \setlength{\fboxsep}{1pt}\fbox{{100.0}} & 98.1 & 116 
& \setlength{\fboxsep}{1pt}\fbox{{3.38}} & 21.27 & 92.2  \\
\bottomrule
\end{tabular}
}
\end{footnotesize}
\end{center}
\vskip -0.25in
\end{table*}

\begin{table*}[t!]
\setlength{\tabcolsep}{6pt}
\vspace{-5pt}
\caption{Discriminative and generative downstream tasks. \textbf{Best}, \underline{second-best} and \setlength{\fboxsep}{1pt}\fbox{{best non-streaming}} results are highlighted.}
\label{tab:downstream}
\begin{center}
\begin{scriptsize}
\resizebox{0.95\textwidth}{!}{%
\begin{tabular}{lc|c|c|c|c|c|ccc|ccc}
\toprule
\multirow{3}{*}{\textbf{Codec}} & \multirow{3}{*}{\makecell{\textbf{Bitrate} \\ \textbf{(kbps)} } $\downarrow$} 
& \multicolumn{1}{c|}{\textbf{ASR}} & \multicolumn{1}{c|}{\textbf{SI}} & \multicolumn{1}{c|}{\textbf{SER}} & \multicolumn{1}{c|}{\textbf{KS}} & \multicolumn{1}{c|}{\textbf{IC}} 
& \multicolumn{3}{c|}{\textbf{SE}} & \multicolumn{3}{c}{\textbf{SS}} \\
\cmidrule{3-13}
& & \textbf{WER} $\downarrow$ & \textbf{ER} $\downarrow$ & \textbf{ER} $\downarrow$ & \textbf{ER} $\downarrow$ & \textbf{ER} $\downarrow$
& \textbf{DNSMOS} $\uparrow$ & \textbf{dWER} $\downarrow$ & \makecell{\textbf{Sim}} $\uparrow$
& \textbf{DNSMOS} $\uparrow$ & \textbf{dWER} $\downarrow$ & \makecell{\textbf{Sim}} $\uparrow$ \\
\midrule
\multicolumn{1}{l}{Reference} & --- 
& --- & --- & --- & --- & --- 
& 3.56 & 0.00 & 100.0 & 3.77 & 0.00 & 100.0 \\

\multicolumn{1}{l}{EnCodec} & 1.50 
& 28.55 & 3.25 & 41.94 & 96.16 & 49.79 
& 3.13 & 37.31 & 85.6 & 3.11 & \underline{77.61} & 87.4 \\

\multicolumn{1}{l}{AudioDec} & 1.60 
& 29.21 & \textbf{1.69} & 45.85 & 25.30 & 46.77 
& 2.96 & 61.11 & 84.3 & 2.97 & {88.59} & 84.0 \\

\multicolumn{1}{l}{HILCodec} & 1.50 
& 29.89 & \underline{1.98} & 51.61 & {15.17} & 53.69 
& 3.32 & 41.33 & \textbf{90.2} & 3.35 & 78.43 & 86.9 \\

\multicolumn{1}{l}{Mimi5} & 0.69 
& 22.56 & 3.92 & 37.10 & 5.99 & {35.03} 
& 3.18 & 52.11 & 86.3 & 3.37 & 87.63 & {88.8} \\

\multicolumn{1}{l}{Mimi6} & 0.83 
& 22.56 & {3.13} & {35.71} & {5.81} & 35.74 
& 3.14 & 55.99 & 86.7 & 3.32 & 86.49 & 88.9 \\

\multicolumn{1}{l}{PAST} & 1.00 
& \textbf{10.74} & 3.43 & {36.41} & 6.41 & {31.66} 
& 3.15 & \textbf{18.19} & 77.9 & 3.15 & 85.61 & 80.3 \\

\midrule

\multicolumn{1}{l}{\textbf{FocalCodec-S@50-2k}} & \textbf{0.55} 
& 16.87 & 2.66 & 37.10 & \textbf{5.50} & 30.93 
& \underline{3.54} & 21.24 & 85.2 & \underline{3.68} & 83.41 & 89.8 \\

\multicolumn{1}{l}{\textbf{FocalCodec-S@50-4k}} & \underline{0.60} 
& 17.21 & 2.70 & \textbf{33.18} & 5.85 & \underline{29.74} 
& \underline{3.54} & {20.21} & \underline{88.7} & \textbf{3.69} & {80.21} & \underline{90.0} \\

\multicolumn{1}{l}{\textbf{FocalCodec-S@50-65k}} & 0.80 
& \underline{17.02} & {2.18} & \underline{34.56} & \underline{5.63} & \textbf{29.49} 
& \textbf{3.56} & \underline{19.56} & 87.7 & \underline{3.68} & \textbf{75.43} & \textbf{90.8} \\
\midrule
\multicolumn{1}{l}{FocalCodec@50} & 0.65 
& 15.33 &  \setlength{\fboxsep}{1pt}\fbox{{0.35}} & 34.79 & \setlength{\fboxsep}{1pt}\fbox{{4.23}} & \setlength{\fboxsep}{1pt}\fbox{{24.66}} 
& 3.52 & \setlength{\fboxsep}{1pt}\fbox{{12.35}} & \setlength{\fboxsep}{1pt}\fbox{{90.4}} & \setlength{\fboxsep}{1pt}\fbox{{3.71}} & \setlength{\fboxsep}{1pt}\fbox{{72.61}}  & 89.5 \\
\bottomrule
\end{tabular}
}
\end{scriptsize}
\end{center}
\vspace{-0.25in}
\end{table*}

\subsection{Downstream tasks}
To assess the quality of the learned discrete representations, we evaluate downstream models on discriminative and generative tasks.
For discriminative tasks, we consider automatic speech recognition (\textbf{ASR}), speaker identification (\textbf{SI}), speech emotion recognition (\textbf{SER}), keyword spotting (\textbf{KS}), and intent classification (\textbf{IC}).
For generative tasks, we focus on speech enhancement (\textbf{SE}) and speech separation (\textbf{SS}). We leave autoregressive tasks such as text-to-speech and speech language modeling for future work.
These tasks are aligned with \cite{mousavi2024dasb}, which provides a standardized evaluation suite for audio discrete representations.
As in \cite{dellalibera2025focalcodec}, we adopt shallow LSTM-based probes for discriminative tasks and Conformer-based non-autoregressive models for SE and SS. Specifically, we use LibriSpeech-460 for ASR and SI, IEMOCAP~\cite{busso2008iemocap} for SER, Speech Commands~\cite{warden2018speechcommands} for KS, SLURP~\cite{bastianelli2020slurp} for IC, VoiceBank~\cite{valentinibotinhao2016voicebank} for SE, and Libri2Mix-100~\cite{cosentino2020librimix} for SS.
Unlike \cite{dellalibera2025focalcodec}, where token embeddings were initialized from the codec embedding layer, we instead use the reconstructed features taken after the quantization bottleneck and before the decoder. We find these features consistently improve performance across most tasks and codecs.
We report error rates for discriminative tasks, and standard speech quality metrics for generative tasks, using DNSMOS~\cite{reddy2022dnsmos} instead of UTMOS (which is better suited for noisy data). For further details, we refer to \cite{dellalibera2025focalcodec}.

As shown in \cref{tab:downstream}, across discriminative tasks FocalCodec-Stream consistently ranks among the top-performing codecs despite operating at lower bitrates. For ASR, all variants achieve a WER of roughly 17\%, substantially outperforming all codecs except PAST, which benefits from having been specifically fine-tuned on this task-dataset combination during training. On SI, the 65k variant obtains the best error rate, outperforming Mimi and PAST, and falling only slightly behind AudioDec and HILCodec, confirming the intuition that preserving speaker information is easier at larger codebook sizes. For SER, performance is strong, with the 4k variant achieving the best results and even marginally surpassing the non-streaming model. On KS and IC, FocalCodec-Stream models outperform all prior baselines. Finally, we note that in the KS task EnCodec performs particularly poorly, likely due to overfitting to commands present in the training set but absent from the test set.

For generative tasks, the benefits of the streaming distillation framework are even more evident. On SE, all three FocalCodec-Stream variants achieve high DNSMOS and competitive speaker similarity. Importantly, their dWER values surpass those of all other baselines except PAST, which is slightly ahead, but conversely shows much lower DNSMOS and speaker similarity. A similar trend is observed for SS, where the 65k variant delivers strong performance across all metrics, substantially outperforming all baselines and rivaling the quality of the non-streaming model.

Finally, it is worth noting that a performance gap with the full-context FocalCodec@50 baseline remains, especially at lower bitrates, which is expected given the stricter constraints of real-time streaming. Nevertheless, the results demonstrate the effectiveness of our codec, which achieves a favorable balance among reconstruction quality, downstream task performance, latency, and efficiency. In particular, the 2k variant is of special interest, as its compact codebook makes it well-suited for autoregressive modeling.

\subsection{Ablation studies}
We conduct ablations to assess the impact of the refiner and stage-4 fine-tuning on resynthesis quality using LibriSpeech \texttt{test-clean}. As shown in \cref{tab:ablations}, the full model achieves the best scores across UTMOS, dWER, and similarity. Removing the refiner consistently degrades performance, reflecting its role in mitigating the distribution shift between causal and full-context features. Omitting the final fine-tuning stage has an even stronger effect, with higher dWER and reduced UTMOS and similarity. These results emphasize the importance of the progressive multi-stage framework for preserving quality under low-latency constraints.

\begin{table}[t!]
\setlength{\tabcolsep}{8pt}
\vspace{-10pt}
\caption{Ablation studies for FocalCodec-S@50-4k.}
\vspace{0.1cm}
\label{tab:ablations}
\centering
\resizebox{0.40\textwidth}{!}{%
\begin{tabular}{lccc}
\toprule
\textbf{Configuration} & \makecell{\textbf{UTMOS} $\uparrow$} & \makecell{\textbf{dWER} $\downarrow$} & \makecell{\textbf{Sim} $\uparrow$} \\
\cmidrule{1-4}
Proposed & \textbf{3.87} & \textbf{4.39} & \textbf{96.3} \\
\, w/o refiner & 3.84 & 4.65 & 96.1 \\
\, w/o stage-4 & 3.78 & 5.05 & 95.8 \\
\bottomrule
\end{tabular}
}
\vspace{-12pt}
\end{table}

\section{Conclusion}
\label{sec:conclusions}
We introduced a streamable extension of FocalCodec that combines a causal single-codebook design with a multi-stage distillation strategy. The proposed codec operates at bitrates as low as 0.55 - 0.80 kbps with an 80 ms theoretical latency, achieving strong reconstruction quality while preserving both semantic and acoustic information, and consistently outperforms popular streamable baselines at comparable bitrates. Future work will focus on scaling to larger datasets, exploring autoregressive modeling, and further optimizing the system for resource-constrained settings.

\bibliographystyle{IEEEbib}
\bibliography{refs}

\end{document}